\documentclass[useAMS,usenatbib]{mn2e}

\usepackage{graphicx}
\usepackage{gensymb}
\usepackage{natbib}
\usepackage{amssymb}
\usepackage{upgreek}
\usepackage{array}
\usepackage{dcolumn}

\newcommand{\grd}{GRD }
\newcommand{\grdd}{GRD}
\newcommand{\ngd}{NGRD }
\newcommand{\ngdd}{NGRD}

\title[On the difference between $\gamma$-ray-loud and quiet pulsars]{On the difference between $\gamma$-ray-detected and non-$\gamma$-ray-detected pulsars}
\author[S. C. Rookyard, P. Weltevrede, S. Johnston and M. Kerr]{S. C. Rookyard$^{1}$, P. Weltevrede$^{1}$\thanks{E-mail: patrick.weltevrede@manchester.ac.uk}, S. Johnston$^{2}$ and M. Kerr$^{2}$\\
$^{1}$Jodrell Bank Centre for Astrophysics, School of Physics and Astronomy, University of Manchester, Manchester M13 9PL, UK\\
$^{2}$CSIRO Astronomy and Space Science, Australia Telescope National Facility, Epping NSW 1710, Australia}
\begin{document}

\date{Accepted 2016 September 19. Received 2016 September 17; in original form 2016 July 11}

\pagerange{\pageref{firstpage}--\pageref{lastpage}} \pubyear{2002}

\maketitle

\label{firstpage}

\begin{abstract}
We compare radio profile widths of young, energetic $\gamma$-ray-detected and non-$\gamma$-ray-detected pulsars. We find that the latter typically have wider radio profiles, with the boundary between the two samples exhibiting a dependence on the rate of rotational energy loss. We also find that within the sample of $\gamma$-ray-detected pulsars, radio profile width is correlated with both the separation of the main $\gamma$-ray peaks and the presence of narrow $\gamma$-ray components. These findings lead us to propose that these pulsars form a single population where the main factors determining $\gamma$ ray detectability are the rate of rotational energy loss and the proximity of the line of sight to the rotation axis. The expected magnetic inclination angle distribution will be different for radio pulsars with and without detectable $\gamma$ rays, naturally leading to the observed differences. Our results also suggest that the geometry of existing radio and outer-magnetosphere $\gamma$-ray emission models are at least qualitatively realistic, implying that information about the viewing geometry can be extracted from profile properties of pulsars.

\end{abstract}


\section{Introduction}
\label{SectIntro}

The emission of rotationally-powered pulsars is believed to ultimately be derived from the loss of rotational energy as the neutron star's rotational period increases. Observationally it is found that the pulsars which emit significant amounts of high energy (i.e. $\gamma$ ray) emission have high rates of rotational energy loss, $\dot{E}$ (e.g., \citealt{sgc+08a}). However, there are also many high-$\dot{E}$ pulsars which are as yet undetected in $\gamma$ rays \citep{aaa+13}. This raises the question of what physical factors other than $\dot{E}$ and distance may determine whether a pulsar is detectable in $\gamma$ rays.  

\cite{rwj15a} analysed the radio emission of a sample of $\gamma$-ray-detected (\grdd) energetic (high $\dot{E}$) pulsars detected by the \emph{Fermi} satellite's Large Area Telescope which are included in the Parkes telescope's \emph{Fermi} timing programme \citep{wjm+10}. Subsequent analysis of the magnetic inclination angle ($\alpha$) distribution \citep{rwj15b} revealed two alternative possibilities: that pulsars are born with a specific non-random $\alpha$ distribution, or that the radio beamsize depends on $\alpha$. In this paper we compare the radio profile morphologies of high-$\dot{E}$ \grd pulsars with a sample of similarly energetic, but non-$\gamma$-ray-detected (\ngdd), pulsars with the aim of investigating what, apart from $\dot{E}$, affects $\gamma$-ray detectability.

\begin{figure} 
\centering 
\includegraphics[height=\hsize,angle=0]{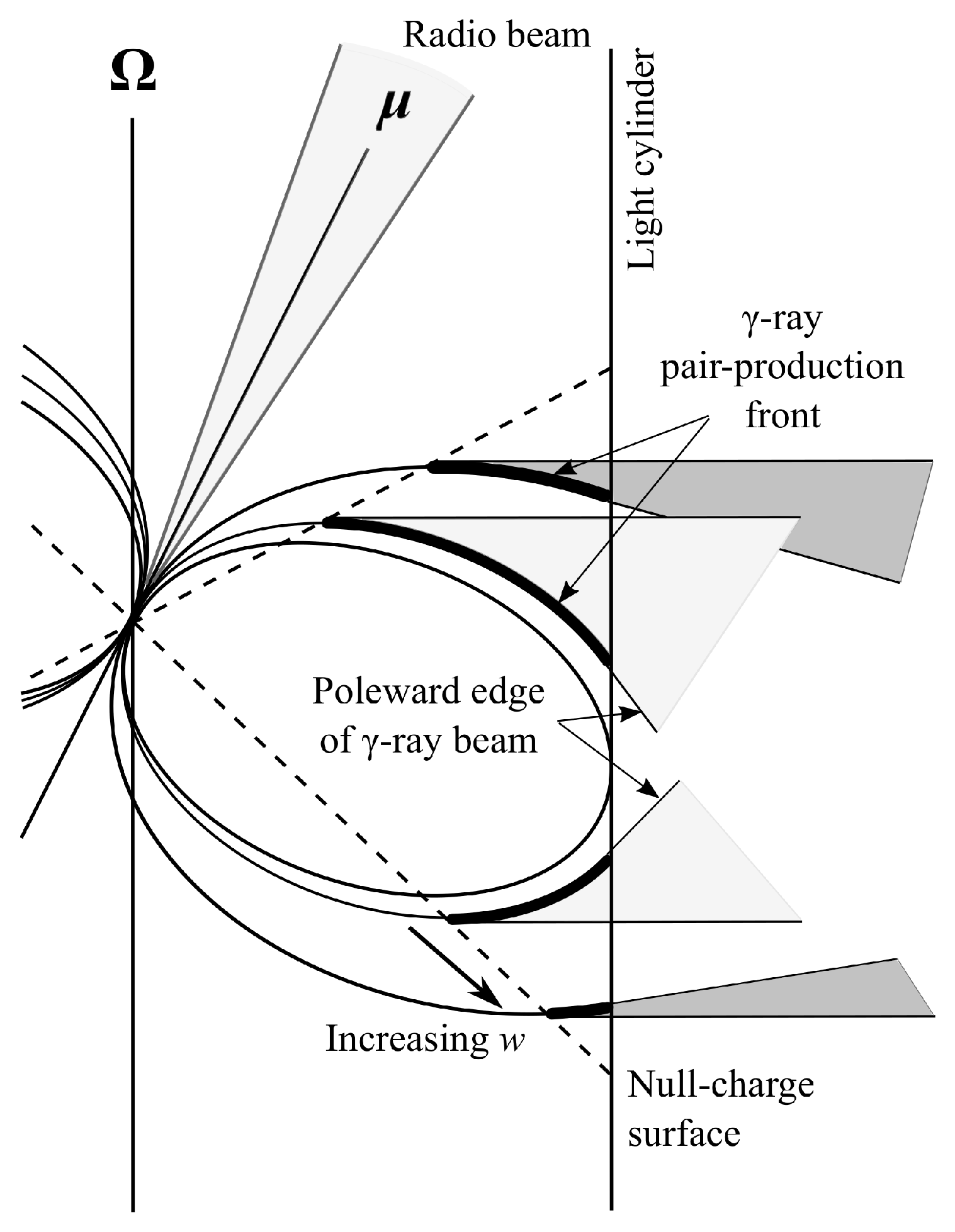} 
\caption{\label{FigBeamGeometry}Geometry of the emission regions, using the outer gap model for the $\gamma$-ray emission as an example. The production of the $\gamma$-ray emission takes place along the pair-production front (ppf; bold lines), the position of which relative to the last open field lines is parameterised by $w$ (see $\S$~\ref{SectLargeAlpha}). The $\gamma$ rays are beamed along tangents to the ppf, resulting in a beam of emission associated with each pole (light and dark grey shaded regions for small and large $w$ respectively). The radio emission is generated on open field lines close to the stellar surface and is centred on the magnetic axis (${\bf \mu}$). The rotation axis (${\bf \Omega}$), the light cylinder (at which the corotation velocity equals the speed of light) and the null-charge surface \citep{gj69} are also shown. }
\end{figure}

In this paper we assume the same generic radio emission geometry as \cite{rwj15a,rwj15b}. The model (illustrated in Fig.~\ref{FigBeamGeometry}) takes the radio emission to originate close to the neutron star and to be confined to
the open field line region, those magnetic dipole field lines that penetrate the light cylinder. This results in a conal beam, the opening angle of which will be sensitive to the altitude at which the emission is generated. There is some uncertainty regarding the precise structure of the radio emission within the beam. Various models have been proposed, such as the `core-cone' model \citep{ran90,ran93}, or the `patchy beam' model \citep{lm88}. However, the generic radio beam model used here is largely insensitive to the structure of the beam provided the emission extends approximately to the edges of the open field line region. An alternative to these circular beams is the `fan-beam' model \citep{mic87a,drd10,wpz+14}, in which the emission is generated in elongated subbeams arranged in a spoke-like pattern centred at the magnetic axis. We will discuss this model separately where its predictions differ from those of the core-cone and patchy beam models.

In contrast to the radio emission, the $\gamma$ rays are believed to be generated at much higher (and over a much more extended range of) altitudes, close to the light cylinder (see Fig.~\ref{FigBeamGeometry}). In all currently favoured $\gamma$-ray emission models, for example the outer gap or two-pole caustic models \citep{chr86a, dr03}, the $\gamma$ rays are generated along a pair-production front (ppf) between some inner limit, such as the null-charge surface \citep{gj69}, and some outer limit. The photons are emitted along tangents to the field lines similarly to the radio emission. Detailed modelling of high-altitude $\gamma$-ray emission has been undertaken by, for example, \cite{wrw+09} and \cite{rw10}, which we use as a basis for our considerations of the $\gamma$-ray beam shape. In this paper we presume the ppf to have zero thickness for simplicity. We also note that some models of the magnetosphere (e.g., the force-free modelling of \citealt{spi06} and \citealt{cra14}) suggest that the $\gamma$-ray beam has a more complicated shape than described here. However, we believe that neither of these assumptions will make a significant qualitative difference to the conclusions of this paper.   

The structure of this paper is as follows: In $\S$~\ref{SectResults} we define the \grd and \ngd samples and present correlations of the radio profile width with both $\gamma$-ray detectability and $\gamma$-ray light curve morphology. In $\S$~\ref{SectPossibleExplanations} we discuss possible physical explanations for these correlations. In $\S$~\ref{SectWidthVersusEdot} we discuss the observed dependence of the radio profile width on $\dot{E}$ and consider the implications. Finally, in $\S$~\ref{SectConclusions} we summarise our findings and draw some conclusions about pulsar emission geometry.

\section{Observational results}
\label{SectResults}

\subsection{The samples}
\label{SectSamples}

We require two samples of young, high-$\dot{E}$ \ngd and \grd pulsars for which the radio emission can be compared. We define the samples as those pulsars which have been detected at radio frequencies, excluding millisecond pulsars ($P < 0.01 \mathrm{s}$), with $\dot{E} = 3.95 \times 10^{31} \mathrm{\ erg\ s}^{-1} (\dot{P}/10^{-15}) (P/\mathrm{sec})^{-3} \geq 10^{35} \mathrm{\ erg\ s}^{-1}$ (for rotational period $P$ and its time derivative $\dot{P}$). This cutoff in $\dot{E}$ was chosen as it represents a marked decrease in the fraction of the population with $\gamma$-ray detections, as shown by Fig.~3 in \cite{lsg+15}. In this paper we will consider radio profile widths, which we measure at a single frequency of $\sim1400$~MHz in order to avoid the effects of radius-to-frequency mapping \citep{kom70}. In addition we exclude pulsars which exhibit the effects of interstellar scattering and those for which the only profiles at $\sim~1400$~MHz available in the literature were of insufficient quality for reliable width estimation. These criteria result in two samples with 35 objects in each case and similar distributions in the $P$-$\dot{P}$ diagram. Table~\ref{TabParameters} lists the relevant properties for each pulsar, including the $\gamma$-ray detectability.

The \emph{Fermi} satellite has been used to search for pulsed $\gamma$-rays from nearly all ($\sim~85\%$) pulsars with $\dot{E} > 10^{34} \mathrm{\ erg\ s^{-1}}$, meaning that the designation of a pulsar as \ngd is not due to a lack of observations of that object \citep{sgc+08a, lsg+15}. Those that were excluded are likely to have been those with low radio luminosities which would correspondingly require large amounts of observing time before $\gamma$-ray searches could be performed. However, provided the radio luminosity is not highly dependent on $\alpha$\footnote{Magnetospheric modeling shows that variations in $\alpha$ can alter $\dot{E}$ by up to a factor $\sim$2 (see, e.g., Fig.~3 of \citealt{lst12}), which in general is unlikely to make a difference to whether a pulsar's $\gamma$-ray emission is above or below the detection threshold. }, we expect that excluding these faint pulsars should not have introduced any significant bias into our conclusions.

One further consideration is the distances of the pulsars, which affects the $\gamma$-ray detectability. As expected the \ngd pulsars have slightly greater (dispersion-measure-based) distances (ATNF Pulsar Catalogue; \citealt{mhth05}). However, a Kolmogorov-Smirnov (KS) test reveals that this difference has a significance $< 3\sigma$. None of the conclusions of this paper are critically dependent on this difference in the distance distribution.

\subsection{\grd pulsars have narrower radio profiles}
\label{SectSampleSeparation}

\begin{figure} 
\centering 
\includegraphics[height=\hsize,angle=270]{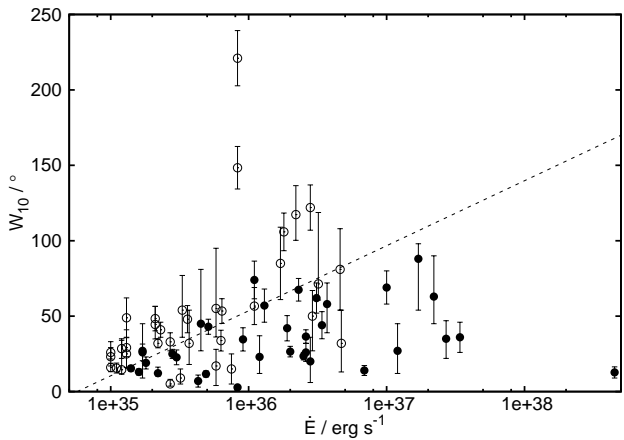} 
\caption{\label{FigWOfEdot}Observed radio profile width (with 3$\sigma$ uncertainty) as a function of $\dot{E}$. Open and filled circles represent \ngd and \grd pulsars respectively. The diagonal dotted line demonstrates the  separation of the samples qualitatively. }
\end{figure}

There is a clear difference in the radio properties of the \grd and \ngd samples. Fig.~\ref{FigWOfEdot} shows $W_{10}$, the pulse width measured at a 10\% intensity level, as a function of $\dot{E}$. The \ngd pulsars (open circles) typically exhibit wider radio profiles than the \grd pulsars (filled circles), although there is overlap between the samples.
The separation of the samples is delineated by the dotted line in the figure, which is intended only to guide the eye. A KS test\footnote{The distribution of vertical offsets from the dotted line was compared for the \ngd and \grd samples. The test statistic  of this test results in a probability of $2.75 \times 10^{-6}$, meaning that the samples are different to a confidence of almost 5$\sigma$. } reveals a significance of almost 5$\sigma$ for this separation.

\subsection{Correlation between radio and $\gamma$-ray light curve morphology}
\label{SectWidthVersusMorphology}

\begin{figure} 
\centering 
\includegraphics[height=\hsize,angle=270]{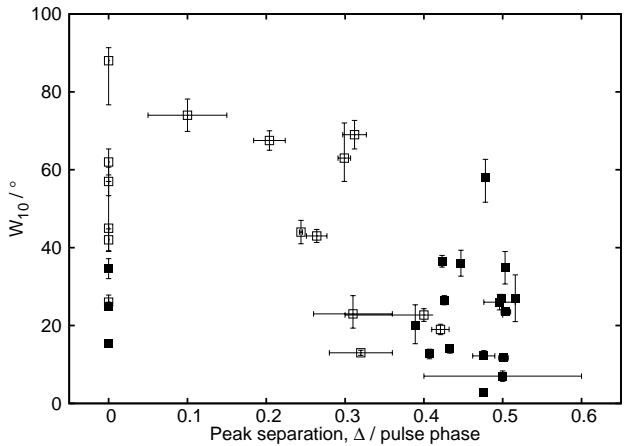} 
\caption{\label{FigWOfDelta}Observed radio profile width versus separation of the two main peaks ($\Delta$) of the $\gamma$-ray light curve. $\Delta$ is expressed as a proportion of the stellar rotation period. Pulsars with $\Delta = 0$ are those with only a single component in the light curve. Open and filled squares represent pulsars with broad and narrow $\gamma$-ray light curve components respectively. The error bars on both parameters represent 1$\sigma$ uncertainties. }
\end{figure}

A correlation between the $\gamma$-ray light curve and radio profile morphologies within the \grd sample has been revealed. One metric of the $\gamma$-ray light curve morphology is $\Delta$, the separation of the two main $\gamma$-ray peaks. Fig.~\ref{FigWOfDelta} shows the radio profile width as a function of $\Delta$ for the \grd pulsars (see Table~\ref{TabParameters} for references). Pulsars without a double-peaked $\gamma$-ray light curve morphology have been assigned $\Delta=0$ in the figure. Ignoring those pulsars, it is clear that pulsars with wide radio profiles have narrower (smaller $\Delta$) $\gamma$-ray light curves. The Spearman rank-ordered correlation coefficient for the pulsars with $\Delta \neq 0$ is $-0.43 ^{+0.13}_{-0.16}$ (3$\sigma$ uncertainties)\footnote{These uncertainties were determined by performing a Monte Carlo analysis assuming Gaussian uncertainties for each pulsar in $W_{10}$ and $\Delta$.}.

In addition to the light curve component separation, the widths of the individual $\gamma$-ray light curve components were also considered. The \grd pulsars are divided into two classes: `narrow-component' (full-width at half maximum\footnote{Note that for the purposes of light curve classification here, $W_{50}$ refers to \emph{component} width, not the overall light curve width.}, $W_{50} < 0.15 P$; filled squares in Fig.~\ref{FigWOfDelta}) and `broad-component' (open squares). A brief discussion of our interpretation of each $\gamma$-ray light curve can be found in the Appendix. The pulsars with narrow radio profiles are predominantly those which exhibit narrow $\gamma$-ray light curve components, whereas pulsars with wider radio profiles tend to have broader $\gamma$-ray components. A KS test between the radio $W_{10}$ values of these two classes yields a probability of $8.45 \times 10^{-4}$, signifying a difference between the two sets of values to a confidence greater than 3$\sigma$. This is another striking model-independent relationship between the radio and $\gamma$-ray profile morphologies.

\begin{table*}
\footnotesize
\centering
\setlength{\extrarowheight}{1 mm}
\caption{\label{TabParameters} Table of the parameters considered in this paper. The columns are, in order, the name of the pulsar, $\dot{E}$, the distance to the pulsar according to the \protect \cite{tc93} Galactic electron density model (ATNF Pulsar Catalogue; \protect \citealt{mhth05}), the measured profile width at the 10\% intensity level, whether the pulsar has been detected in $\gamma$-rays, the $\gamma$-ray peak separation $\Delta$ as a proportion of the pulse period and the classification of the $\gamma$-ray light curve according to the full width at half-maximum of the components. The references for light curve class also apply to $\Delta$. References: [1] \protect \cite{phs+15}; [2] \protect \cite{bkr+13}; [3] \protect \cite{mh99}; [4] \protect \cite{rwj15a}; [5] \protect \cite{roo15}; [6] \protect \cite{hsg+14}; [7] \protect \cite{rkp+11}; [8] Unpublished Parkes Data obtained for the ongoing \emph{Fermi} timing project \protect \citep{wjm+10} at 1369~MHz; [9] \protect \cite{aaa+13}; [10] \protect \cite{kel+09}; [11] Unpublished archival Lovell telescope data at 1532~MHz; [12] \protect \cite{vks+15}; [13] \protect \cite{aaa+10a}; [14] \protect \cite{lx00}; [15] \protect \cite{mhl+02}; [16] \protect \cite{lfl+06}; [17] \protect \cite{lcm13}; [18] \protect \cite{aaa+10c}; [19] \protect \cite{bck+13}; [20] \protect \cite{crr+09}; [21] \protect \cite{hcg+01}; [22] \protect \cite{tpc+11}; [23] \protect \cite{lsg+15}. }

\begin{tabular}{lD{.}{.}{4.1}D{.}{.}{2.2}r@{ $^{+}_{-}$ }rcr@{ $\pm$ }lc}
\hline
PSR & \multicolumn{1}{c}{$\dot{E}$} & \multicolumn{1}{c}{Distance} & \multicolumn{2}{c}{$W_{10} / \degree$} & \grd or & \multicolumn{2}{c}{$\Delta$} & Light curve\\
& \multicolumn{1}{c}{/ $10^{35} \mathrm{\ erg\ s}^{-1}$} & \multicolumn{1}{c}{/ kpc} & \multicolumn{2}{c}{(3$\sigma$ uncertainty)} & \ngd & \multicolumn{2}{c}{(1$\sigma$ uncertainty)} & class \\
\hline

J0117+5914 & 2.2 & 2.14 & 32.0 & $^{3.0}_{3.0}$ $^{[1]}$ & NGRD & \multicolumn{2}{c}{---} & --- \\ 
J0205+6449 & 270.0 & 7.53 & 35.0 & $^{12.0}_{13.0}$ $^{[2]}$ & GRD & 0.503 & 0.004 & narrow $^{[9]}$\\ 
J0534+2200 & 4500.0 & 2.49 & 12.7 & $^{3.7}_{3.7}$ $^{[3]}$ & GRD & 0.407 & 0.001 & narrow $^{[9]}$\\ 
J0631+1036 & 1.7 & 6.54 & 26.0 & $^{5.5}_{5.5}$ $^{[4]}$ & GRD & \multicolumn{2}{c}{---} & broad $^{[9]}$\\ 
J0729$-$1448 & 2.8 & 4.37 & 25.0 & $^{3.0}_{3.0}$ $^{[4]}$ & GRD & \multicolumn{2}{c}{---} & narrow $^{[9]}$\\ 
J0742$-$2822 & 1.4 & 1.89 & 15.4 & $^{1.8}_{1.8}$ $^{[4]}$ & GRD & \multicolumn{2}{c}{---} & narrow $^{[9]}$\\ 
J0835$-$4510 & 69.0 & 0.61 & 14.0 & $^{3.3}_{3.3}$ $^{[4]}$ & GRD & 0.433 & 0.001 & narrow $^{[9]}$\\ 
J0855$-$4644 & 11.0 & 9.90 & 56.7 & $^{12.3}_{12.3}$ $^{[5]}$ & NGRD & \multicolumn{2}{c}{---} & --- \\ 
J0908$-$4913 & 4.9 & 6.66 & 11.7 & $^{2.0}_{2.0}$ $^{[4]}$ & GRD & 0.501 & 0.006 & narrow $^{[9]}$\\ 
J0940$-$5428 & 19.0 & 4.27 & 42.0 & $^{8.4}_{8.4}$ $^{[4]}$ & GRD & \multicolumn{2}{c}{---} & broad $^{[9]}$\\ 
J1015$-$5719 & 8.3 & 4.87 & 148.4 & $^{14.1}_{14.1}$ $^{[5]}$ & NGRD & \multicolumn{2}{c}{---} & --- \\ 
J1016$-$5857 & 26.0 & 9.31 & 36.5 & $^{4.5}_{4.5}$ $^{[4]}$ & GRD & 0.423 & 0.004 & narrow $^{[9]}$\\ 
J1028$-$5819 & 8.3 & 2.76 & 2.8 & $^{0.4}_{0.4}$ $^{[4]}$ & GRD & 0.475 & 0.001 & narrow $^{[9]}$\\ 
J1048$-$5832 & 20.0 & 2.98 & 26.5 & $^{3.5}_{3.5}$ $^{[4]}$ & GRD & 0.426 & 0.001 & narrow $^{[9]}$\\ 
J1052$-$5954 & 1.3 & 13.55 & 48.9 & $^{13.2}_{13.2}$ $^{[5]}$ & NGRD & \multicolumn{2}{c}{---} & --- \\ 
J1055$-$6028 & 12.0 & 0.45 & 23.0 & $^{14.0}_{11.0}$ $^{[6]}$ & GRD & 0.31 & 0.05 & broad $^{[6]}$\\ 
J1105$-$6107 & 25.0 & 7.07 & 23.5 & $^{2.5}_{2.5}$ $^{[4]}$ & GRD & 0.504 & 0.006 & narrow $^{[9]}$\\ 
J1119$-$6127 & 23.0 & 8.40 & 67.5 & $^{7.5}_{7.5}$ $^{[4]}$ & GRD & 0.204 & 0.02 & broad $^{[9]}$\\ 
J1124$-$5916 & 120.0 & 10.93 & 27.0 & $^{18.0}_{15.0}$ $^{[7]}$ & GRD & 0.499 & 0.004 & narrow $^{[9]}$\\ 
J1301$-$6305 & 17.0 & 15.84 & 85.0 & $^{24.0}_{24.0}$ $^{[5]}$ & NGRD & \multicolumn{2}{c}{---} & --- \\ 
J1302$-$6350 & 8.3 & 4.60 & 221.0 & $^{18.3}_{18.3}$ $^{[5]}$ & NGRD & \multicolumn{2}{c}{---} & --- \\ 
J1357$-$6429 & 31.0 & 4.09 & 62.0 & $^{10.0}_{10.0}$ $^{[4]}$ & GRD & \multicolumn{2}{c}{---} & broad $^{[9]}$\\ 
J1359$-$6038 & 1.2 & 5.90 & 14.5 & $^{19.5}_{3.0}$ $^{[5]}$ & NGRD & \multicolumn{2}{c}{---} & --- \\ 
J1412$-$6145 & 1.2 & 9.32 & 28.5 & $^{6.5}_{6.5}$ $^{[5]}$ & NGRD & \multicolumn{2}{c}{---} & --- \\ 
J1420$-$6048 & 100.0 & 7.65 & 69.0 & $^{11.0}_{11.0}$ $^{[4]}$ & GRD & 0.312 & 0.015 & broad $^{[9]}$\\ 
J1509$-$5850 & 5.1 & 3.85 & 43.0 & $^{5.0}_{5.0}$ $^{[4]}$ & GRD & 0.264 & 0.013 & broad $^{[9]}$\\ 
J1512$-$5759 & 1.3 & 12.70 & 29.1 & $^{11.9}_{5.9}$ $^{[5]}$ & NGRD & \multicolumn{2}{c}{---} & --- \\ 
J1513$-$5908 & 170.0 & 5.79 & 88.0 & $^{10.0}_{34.0}$ $^{[4]}$ & GRD & \multicolumn{2}{c}{---} & broad $^{[9]}$\\ 
J1524$-$5625 & 32.0 & 3.84 & 71.4 & $^{47.3}_{10.7}$ $^{[8]}$ & NGRD & \multicolumn{2}{c}{---} & --- \\ 
J1531$-$5610 & 9.1 & 3.10 & 34.6 & $^{7.7}_{7.7}$ $^{[4]}$ & GRD & \multicolumn{2}{c}{---} & narrow $^{[9]}$\\ 
J1541$-$5535 & 1.1 & 7.46 & 15.6 & $^{3.3}_{3.3}$ $^{[5]}$ & NGRD & \multicolumn{2}{c}{---} & --- \\ 
J1601$-$5335 & 1.0 & 4.04 & 16.0 & $^{2.8}_{2.8}$ $^{[5]}$ & NGRD & \multicolumn{2}{c}{---} & --- \\ 
J1636$-$4440 & 2.1 & 9.33 & 48.4 & $^{8.1}_{8.1}$ $^{[5]}$ & NGRD & \multicolumn{2}{c}{---} & --- \\ 
J1637$-$4642 & 6.4 & 5.77 & 53.4 & $^{8.2}_{8.2}$ $^{[5]}$ & NGRD & \multicolumn{2}{c}{---} & --- \\ 
J1646$-$4346 & 3.6 & 6.86 & 48.0 & $^{9.0}_{9.0}$ $^{[8]}$ & NGRD & \multicolumn{2}{c}{---} & --- \\ 
J1702$-$4310 & 6.3 & 5.44 & 33.8 & $^{6.9}_{6.9}$ $^{[5]}$ & NGRD & \multicolumn{2}{c}{---} & --- \\ 
J1709$-$4429 & 34.0 & 1.82 & 44.0 & $^{9.0}_{9.0}$ $^{[4]}$ & GRD & 0.244 & 0.002 & broad $^{[9]}$\\ 
J1718$-$3825 & 13.0 & 4.24 & 57.0 & $^{11.0}_{11.0}$ $^{[4]}$ & GRD & \multicolumn{2}{c}{---} & broad $^{[9]}$\\ 
J1739$-$3023 & 3.0 & 3.41 & 22.7 & $^{4.9}_{4.9}$ $^{[5]}$ & GRD & 0.4 & 0.1 & broad $^{[23]}$\\ 
J1740+1000 & 2.3 & 1.36 & 41.0 & $^{5.0}_{5.0}$ $^{[8]}$ & NGRD & \multicolumn{2}{c}{---} & --- \\ 

\hline
\multicolumn{9}{r}{continued...}

\end{tabular}
\end{table*}

\addtocounter{table}{-1}
\begin{table*}
\footnotesize
\centering
\setlength{\extrarowheight}{1 mm}
%\captionsetup{justification=centering}
\caption{ -- continued }

\begin{tabular}{lD{.}{.}{4.1}D{.}{.}{2.2}r@{ $^{+}_{-}$ }rcr@{ $\pm$ }lc}
\hline
PSR & \multicolumn{1}{c}{$\dot{E}$} & \multicolumn{1}{c}{Distance} & \multicolumn{2}{c}{$W_{10} / \degree$} & \grd or & \multicolumn{2}{c}{$\Delta$} & Light curve\\
& \multicolumn{1}{c}{/ $10^{35} \mathrm{\ erg\ s}^{-1}$} & \multicolumn{1}{c}{/ kpc} & \multicolumn{2}{c}{(3$\sigma$ uncertainty)} & \ngd & \multicolumn{2}{c}{(1$\sigma$ uncertainty)} & class \\
\hline
J1801$-$2451 & 26.0 & 4.61 & 26.0 & $^{6.0}_{6.0}$ $^{[4]}$ & GRD & 0.496 & 0.02 & narrow $^{[9]}$\\ 
J1803$-$2137 & 22.0 & 3.94 & 117.3 & $^{19.2}_{17.0}$ $^{[5]}$ & NGRD & \multicolumn{2}{c}{---} & --- \\ 
J1809$-$1917 & 18.0 & 3.71 & 105.9 & $^{12.5}_{12.5}$ $^{[5]}$ & NGRD & \multicolumn{2}{c}{---} & --- \\ 
J1826$-$1334 & 28.0 & 4.12 & 122.0 & $^{15.0}_{15.0}$ $^{[8]}$ & NGRD & \multicolumn{2}{c}{---} & --- \\ 
J1831$-$0952 & 11.0 & 4.33 & 74.0 & $^{12.5}_{12.5}$ $^{[5]}$ & GRD & 0.10 & 0.05 & broad $^{[23]}$\\ 
J1833$-$0827 & 5.8 & 5.67 & 55.1 & $^{39.9}_{18.9}$ $^{[5]}$ & NGRD & \multicolumn{2}{c}{---} & --- \\ 
J1833$-$1034 & 340.0 & 3.74 & 36.0 & $^{10.0}_{10.0}$ $^{[9]}$ & GRD & 0.447 & 0.004 & narrow $^{[9]}$\\ 
J1835$-$1106 & 1.8 & 3.08 & 19.0 & $^{4.0}_{4.0}$ $^{[4]}$ & GRD & 0.421 & 0.011 & broad $^{[9]}$\\ 
J1838$-$0549 & 1.0 & 4.73 & 26.0 & $^{7.1}_{4.4}$ $^{[5]}$ & NGRD & \multicolumn{2}{c}{---} & --- \\ 
J1841$-$0345 & 2.7 & 4.15 & 33.0 & $^{6.0}_{6.0}$ $^{[8]}$ & NGRD & \multicolumn{2}{c}{---} & --- \\ 
J1841$-$0524 & 1.0 & 4.89 & 23.1 & $^{6.1}_{6.1}$ $^{[5]}$ & NGRD & \multicolumn{2}{c}{---} & --- \\ 
J1850$-$0026 & 3.3 & 10.69 & 54.0 & $^{23.0}_{18.0}$ $^{[10]}$ & NGRD & \multicolumn{2}{c}{---} & --- \\ 
J1853$-$0004 & 2.1 & 6.58 & 44.5 & $^{12.0}_{9.6}$ $^{[5]}$ & NGRD & \multicolumn{2}{c}{---} & --- \\ 
J1856+0113 & 4.3 & 2.78 & 7.0 & $^{4.0}_{4.0}$ $^{[11]}$ & GRD & 0.5 & 0.1 & narrow $^{[23]}$\\ 
J1856+0245 & 46.0 & 10.29 & 81.0 & $^{27.0}_{27.0}$ $^{[11]}$ & NGRD & \multicolumn{2}{c}{---} & --- \\ 
J1857+0143 & 4.5 & 5.18 & 45.0 & $^{36.0}_{18.0}$ $^{[11]}$ & GRD & \multicolumn{2}{c}{---} & broad $^{[23]}$\\ 
J1906+0746 & 2.7 & 4.53 & 5.1 & $^{3.0}_{1.5}$ $^{[12]}$ & NGRD & \multicolumn{2}{c}{---} & --- \\ 
J1907+0602 & 28.0 & 3.01 & 20.0 & $^{16.0}_{14.0}$ $^{[13]}$ & GRD & 0.389 & 0.004 & narrow $^{[9]}$\\ 
J1907+0918 & 3.2 & 7.68 & 9.0 & $^{6.0}_{4.0}$ $^{[14]}$ & NGRD & \multicolumn{2}{c}{---} & --- \\ 
J1909+0912 & 1.3 & 8.20 & 25.0 & $^{11.0}_{9.0}$ $^{[15]}$ & NGRD & \multicolumn{2}{c}{---} & --- \\ 
J1913+0904 & 1.6 & 3.49 & 13.0 & $^{2.0}_{2.0}$ $^{[16]}$ & GRD & 0.32 & 0.04 & broad $^{[6]}$\\ 
J1913+1011 & 29.0 & 4.48 & 50.0 & $^{17.0}_{23.0}$ $^{[15]}$ & NGRD & \multicolumn{2}{c}{---} & --- \\ 
J1932+2220 & 7.5 & 9.63 & 15.0 & $^{10.0}_{10.0}$ $^{[11]}$ & NGRD & \multicolumn{2}{c}{---} & --- \\ 
J1935+2025 & 47.0 & 8.64 & 32.0 & $^{22.0}_{19.0}$ $^{[17]}$ & NGRD & \multicolumn{2}{c}{---} & --- \\ 
J1938+2213 & 3.7 & 4.54 & 32.0 & $^{22.0}_{14.0}$ $^{[17]}$ & NGRD & \multicolumn{2}{c}{---} & --- \\ 
J1952+3252 & 37.0 & 2.37 & 58.0 & $^{14.0}_{19.0}$ $^{[18]}$ & GRD & 0.478 & 0.003 & narrow $^{[9]}$\\ 
J2004+3429 & 5.8 & 22.18 & 17.0 & $^{11.0}_{13.0}$ $^{[19]}$ & NGRD & \multicolumn{2}{c}{---} & --- \\ 
J2032+4127 & 1.7 & 5.15 & 27.0 & $^{18.0}_{18.0}$ $^{[20]}$ & GRD & 0.516 & 0.001 & narrow $^{[9]}$\\ 
J2229+6114 & 220.0 & 12.71 & 63.0 & $^{27.0}_{18.0}$ $^{[21]}$ & GRD & 0.299 & 0.008 & broad $^{[9]}$\\ 
J2240+5832 & 2.2 & 14.85 & 12.2 & $^{4.0}_{1.9}$ $^{[22]}$ & GRD & 0.476 & 0.014 & narrow $^{[9]}$\\ 

\hline

\end{tabular}
\end{table*}

\section{Possible interpretations}
\label{SectPossibleExplanations}

A relationship between radio profile width and $\gamma$-ray detectability could in principle be related to the $\gamma$-ray luminosity. However, an examination of the data reveals no obvious relation between $\gamma$-ray luminosity and $W_{10}$ within the \grd sample, suggesting that the transition from \grd to \ngd pulsars in Fig.~\ref{FigWOfEdot} is not due to a gradual decrease in the $\gamma$-ray emissivity. Additionally, as noted in $\S$~\ref{SectSamples}, magnetospheric modeling suggests that the geometry of the pulsar has only a relatively small effect on $\dot{E}$ (and hence $\gamma$-ray luminosity). In this section we therefore explore four geometrical scenarios which could explain why \ngd pulsars have wider radio profiles. We find that a difference in magnetic inclination ($\alpha$; the angle between the rotation and magnetic axes) between the samples ($\S$~\ref{SectLargeAlpha}) is the most plausible and is almost certainly present. We find that the other scenarios are either implausible or would only operate in conjunction with a systematic difference in $\alpha$.

\subsection{Lower magnetic inclinations for \ngd pulsars}
\label{SectLargeAlpha}

A small $\alpha$ will result in large profile widths (e.g., \citealt{ggr84}). Here we consider the scenario that lower $\alpha$ values are the cause of the larger profile widths in the \ngd sample. As will be argued below, such a relation between $\alpha$ and $\gamma$-ray detectability can be expected from $\gamma$-ray emission modeling, as can the $\dot{E}$-dependence of the boundary between the two samples (dashed line in Fig.~\ref{FigWOfEdot}).

The ppf, along which the $\gamma$-ray emission is generated, lies along a particular field line inside the open field line region (see Fig.~\ref{FigBeamGeometry}). The $\gamma$ rays will then be visible along tangents to this field line, which make a range of angles $\xi$ with the rotation axis. If the ppf extends from the null-charge surface up to higher altitudes, the emission of a given magnetic pole can be expected to be confined to a wedge (mid and dark grey regions in Fig.~\ref{FigBeamGeometry}) extending from $\xi = 90\degree$ (rotational equator) up to a poleward edge of the beam defined by the tangent at the highest emission altitude. Hence, $\gamma$ rays will only be observable for a given pulsar if the angle $\zeta$ between the rotation axis and the line of sight is sufficiently large that the line of sight lies within the wedge of $\gamma$ rays. Considering that $\zeta \approx \alpha$ for radio-detected pulsars with relatively narrow beams, this explains how small $\alpha$ (and hence larger $W_{10}$) values could result in a \ngd pulsar since the line of sight is more likely to miss the $\gamma$-ray beam. Some overlap in $W_{10}$ between the samples at a given $\dot{E}$ can be expected since other parameters also affect the width of the radio profile, such as the radio emission height (hence beamsize) or the impact parameter of the line of sight to the magnetic axis. Such an overlap is indeed observed in Fig.~\ref{FigWOfEdot}.

The position of the ppf is described by a parameter $w$, which is defined as the distance between the foot of the last open field line at the stellar surface and the foot of the field line containing the ppf, as a proportion of the radius of the open field line region \citep{rom96a}. When $w$ is larger the emitting field line will be closer to the magnetic pole and so, assuming a dipolar field, will curve less before it reaches the outer limit of the ppf. Hence, the $\gamma$-ray beam will be confined closer to $\xi = 90\degree$ (see Fig.~\ref{FigBeamGeometry}). This is shown for the outer gap model in Fig.~4 of \cite{wrw+09}, in which it can be seen that when $w$ is larger their simulations generate $\gamma$-ray-detectable pulsars over a more restricted range with both $\alpha$ and $\zeta$ relatively close to $90\degree$. Using the relation $w \propto \dot{E}^{-1/2}$ (see \citealt{wrw+09, mh03}) we find that higher-$\dot{E}$ pulsars should have smaller $w$ and therefore the $\gamma$ rays should be less confined. For the line of sight to miss the $\gamma$-ray beam then requires $\zeta$ to be smaller. Hence for a radio-detected pulsar with $\alpha \approx \zeta$, smaller $\alpha$ values resulting in larger $W_{10}$ can be expected for \ngd pulsars. Therefore an $\dot{E}$-dependence of the boundary between \ngd and \grd pulsars in Fig.~\ref{FigWOfEdot} (dotted line) follows naturally from $\gamma$-ray emission models. As will be discussed in $\S$~\ref{SectWidthVersusEdot}, an increasing radio emission height with $\dot{E}$ would also contribute to the $\dot{E}$-dependence of this boundary.  

Additionally, simulations of $\gamma$-ray light curves by, for example, \cite{wrw+09}, according to the two-pole caustic and outer gap models, suggest that the light curve components are narrower and also that $\Delta$ is larger when $\alpha$ is larger. This would explain the associations of narrow-component $\gamma$-ray light curves and larger $\Delta$ values with lower values of $W_{10}$, as can be seen in Fig.~\ref{FigWOfDelta}. It is therefore apparent that a systematic difference in $\alpha$ between the two samples is able to qualitatively explain all the observational results of $\S$~\ref{SectResults}.

\subsection{Larger emission heights for \ngd pulsars}
\label{SectLargeEmissionHeight}

The wide radio profiles in the \ngd sample can in principle be explained if the emission is typically generated higher in the magnetosphere, as a proportion of the light cylinder radius, which would result in a greater beamwidth. Here we discuss the scenario that a higher radio emission height is the dominant physical difference between the two samples, taking the two samples' $\alpha$ and $\zeta$ distributions to be the same.

Such a scenario implies a mechanism whereby the $\gamma$-ray beam moves out of the line of sight as the radio emission height increases. For this we would require that the $\gamma$-ray beam becomes more confined to the rotational equator, in the outer gap model suggestive of a decrease of the upper limit of the $\gamma$-ray emission height. Hence, an increase in radio emission height would have to cause (or at least be correlated with) a decrease in the \emph{highest} altitude of the $\gamma$-ray emission\footnote{In the two-pole caustic model, the poleward edge of the beam could be contributed by either the lowest or highest altitude $\gamma$-ray emission, meaning that an increase in radio emission height would need to increase the lowest altitude or decrease the highest altitude, and in some circumstances may need to do both. }, the point furthest away from the lower altitude radio emission region. There is currently no theoretical explanation for such a link between the radio and $\gamma$-ray emission heights.

Another way to confine the $\gamma$-ray beam to larger $\xi$ is to increase $w$ (see $\S$~\ref{SectLargeAlpha}). Therefore a higher altitude of radio emission would require a larger $w$ at a given $\dot{E}$. It is physically unclear why there should be such a dependence.

Although we cannot exclude a systematic difference in radio emission height between the two samples as a possibility, we find this scenario to be implausible as it depends on ad-hoc unexplained physics. Further to this, whichever cause we consider for the poleward edge of the $\gamma$-ray beam moving towards $\xi = 90\degree$, a pulsar with sufficiently large $\alpha \approx \zeta$ will be detectable in $\gamma$ rays (and similarly pulsars with sufficiently low $\alpha$ will not be detectable in $\gamma$ rays) regardless of the emission height. We are therefore forced to conclude that, even if the extent of the $\gamma$-ray beam is sensitive to the radio emission height, the scenario detailed in $\S$~\ref{SectLargeAlpha} must also operate at the same time.

\subsection{Larger $s$ values for \ngd pulsars}
\label{SectLargeS}

Wider radio profiles can be expected for larger $s$, the ratio of the size of the emission region to that of a dipolar open field line region (see \citealt{rwj15b}). Here we explore the scenario in which the \ngd pulsars have larger $s$, dominating over any other physical effect, as an explanation for their wider profiles.

A larger value of $s$ means that the \emph{effective} last open field line is further from the magnetic pole. It seems reasonable to expect that if the radio emission region expands to include these field lines, then the $\gamma$-ray emission region might also shift to a field line further away from the magnetic pole. As a result, for a given $w$ (and hence $\dot{E}$) the ppf  would be on a field line with greater curvature (see $\S$~\ref{SectLargeAlpha}), {\em increasing} the  $\gamma$-ray beamwidth. Therefore a larger $s$ (wider radio profile) should increase the probability of detecting the pulsar in $\gamma$ rays, opposite to what is observed and making this scenario unlikely.
Furthermore, it should be noted that even if an increase in $s$ could be shown to somehow confine the $\gamma$-ray beam towards the rotational equator, an analogous argument to that in $\S$~\ref{SectLargeEmissionHeight} means that the $\alpha$ dependence described in $\S$~\ref{SectLargeAlpha} must also be present.

\subsection{Difference in the line of sight}
\label{SectDifferentBetas}

\subsubsection{Smaller impact parameters for \ngd pulsars in a circular beam model}
\label{SectSmallBeta}

The typically narrower observed radio profiles of the \grd sample could indicate that the lines of sight for these pulsars mostly only graze the outer part of the radio beam. For this effect to produce the observed difference in $W_{10}$, the average impact parameter $\beta$ of the \grd sample must be a considerable fraction of the beam radius\footnote{For example, approximating the path of the line of sight across the emission region as a straight line, even decreasing the typical profile width by one third requires an impact parameter at least $\sim0.75$ of the beam radius.}, assuming that the radio emission is distributed over a reasonably large proportion of the open field line region. If radius-to-frequency mapping is present, we would then expect $W_{10}$ to be much more sensitive to (frequency-induced) changes in the radio beamwidth for \grd pulsars than for \ngd pulsars. However, inspection of profiles for these pulsars at various observing frequencies in the literature did not reveal any systematic difference in the frequency-dependence of $W_{10}$ between the samples.

\subsubsection{Larger impact parameters for \ngd pulsars in a fan-beam model}
\label{SectLargeBeta}

In the `fan-beam' model \citep{drd10}, and as also shown by \cite{wpz+14}, the wider profiles in the \ngd sample could in theory be explained by larger $\beta$ parameters for those pulsars\footnote{Although it can be seen from the simulated profiles in the latter paper that the effect of $\alpha$ on $W_{10}$ could still be greater.}. If this is the dominant effect, it would imply that pulsars become undetectable in $\gamma$ rays if $\beta$ is too large while the radio emission remains detectable. However, it is unlikely that the $\gamma$-ray beam is narrower than the radio beam since the pulsed $\gamma$-ray emission is observed to cover a larger range of pulse phase for a given \grd pulsar (e.g., \citealt{aaa+13}). Also, such a scenario is incompatible with the population of radio-quiet $\gamma$-ray pulsars (e.g., \citealt{aaa+13}).

\section{Relationship between $W_{10}$ and $\dot{E}$}
\label{SectWidthVersusEdot}

It can be seen from Fig.~\ref{FigWOfEdot} that there is an overall increase of $W_{10}$ with $\dot{E}$, which is observed in both samples. Although narrow radio profiles are found for the full $\dot{E}$ range, the width of the widest profiles detected is larger at higher $\dot{E}$. We consider two possible explanations.

The first possibility is that the $\alpha$ distribution becomes progressively more skewed towards lower values as $\dot{E}$ increases\footnote{It is unlikely that a difference in the $\alpha$ distribution directly causes the observed differences in $\dot{E}$. As discussed in $\S$~\ref{SectPossibleExplanations}, magnetospheric modeling suggests variations in $\alpha$ are insufficient to explain the orders-of-magnitude range of $\dot{E}$ in our sample.}. A potential cause of this would be a gradual orthogonalisation (i.e., $\alpha \rightarrow 90\degree$) of the magnetic axis with the rotation axis over time. Such an orthogonalisation has recently been proposed for the Crab pulsar \citep{lgw+13}. However it was estimated that $\alpha$ increases by $\sim 0.6\degree$ per century for that pulsar. If this can be extrapolated over the lifetime of a pulsar, which is unclear, the observed decrease in $W_{10}$ over several orders of magnitude in $\dot{E}$ would be too gradual to be explained by such a rapid rate of change of $\alpha$. Other proposed timescales for $\alpha$ evolution (e.g., \citealt{wj08b}) exceed the characteristic ages of this sample. Furthermore, orthogonalisation would also be incompatible with the observational evidence for alignment of the axes over time (\citealt{cb86, lm88, tm98, wj08b}).

The more likely explanation is that the radio beamwidth increases with $\dot{E}$, due to an increase in either $s$ or the radio emission height as a proportion of the light cylinder radius. A least-$\chi^2$ linear fit across both samples suggests that $W_{10}$ is typically a factor of $\sim 2$ larger at $\dot{E} = 10^{37} \mathrm{\ erg\ s}^{-1}$ than at $\dot{E} = 10^{35} \mathrm{\ erg\ s}^{-1}$. If we interpret this in terms of emission height this would imply that the emission height as a proportion of the light cylinder radius increases by a factor of $\sim 4$ over this $\dot{E}$ range. Alternatively, the correlation between $W_{10}$ and $\dot{E}$ could arise without variations in the emission height if $s$ is systematically larger when $\dot{E}$ is higher. 

In the literature, \cite{jw06} concluded that the emission height is typically larger in younger pulsars. Also, \cite{rmh10} used the radio detectability of pulsars which are observed in $\gamma$ rays to suggest that the radio and $\gamma$-ray beaming fractions are similar for very high-$\dot{E}$ pulsars ($\dot{E} > 6 \times 10^{36} \mathrm{\ erg\ s}^{-1}$). A larger radio beamwidth, whether due to larger emission heights or $s$, will increase the radio beaming fraction and be consistent with this conclusion.  

The five objects with largest $W_{10}$ in our sample are ``energetic wide beam'' (EWB) pulsars as discussed by \cite{wj08b}. Those authors noted the radio profile morphologies of EWB pulsars are similar to each other and comparable to typical $\gamma$-ray light curves. This is suggestive of emission produced up to a considerable fraction of the light cylinder radius, close to the site of the $\gamma$ ray production, although these objects are not detected in $\gamma$ rays. The radio profile widths of these pulsars are consistent with the overall trend of increasing $W_{10}$ with $\dot{E}$, which could represent additional evidence that this trend is caused by systematically varying emission heights. We note, however, that to explain their non-detectability in $\gamma$ rays we still require $\alpha$ to be relatively low, hence making the low $\alpha$ a contributing factor to their wide radio profiles. 

It is important to consider how the correlation between $W_{10}$ and $\dot{E}$ relates to the conclusions we reached in $\S$~\ref{SectPossibleExplanations}. If pulsars at higher $\dot{E}$ have wider radio beams, the slope of the dotted line in Fig.~\ref{FigWOfEdot} will depend on both the changing extent of the $\gamma$-ray beam with $w$ (as described in $\S$~\ref{SectLargeAlpha}) and the changing radio beamwidth. However, such a variation in beamwidth does not affect the qualitative arguments against a systematic difference in emission heights or $s$ between the \ngd and \grd pulsars (presented in $\S$~\ref{SectLargeEmissionHeight} and $\S$~\ref{SectLargeS}).

\section{Summary and conclusions}
\label{SectConclusions}

In this paper we have compared two samples of young, high-$\dot{E}$, non-$\gamma$-ray-detected (\ngdd) and $\gamma$-ray-detected (\grdd) radio pulsars which are similar in terms of their spin properties. The most striking difference found between the two samples is a pronounced separation in ($W_{10}$, $\dot{E}$) space, such that \ngd pulsars have wider radio profiles at a given $\dot{E}$. Given that both $W_{10}$ and $\dot{E}$ are directly observable parameters, this result is model-independent. We also report correlations between radio profile width and the widths and peak separations of the $\gamma$-ray light curve components, which is the first time a direct link between radio and $\gamma$-ray profile morphologies has been reported. These results imply some link between the radio and $\gamma$-ray emission mechanisms. These features are not significantly affected by the distances of the pulsars and as such are robust clues to the differences between \ngd and \grd pulsars. For both samples we find that $W_{10}$ increases with $\dot{E}$. It is also interesting to note that according to Fig.~\ref{FigWOfEdot} nearly all pulsars with $10^{34} \mathrm{\ erg\ s^{-1}} < \dot{E} < 10^{35} \mathrm{\ erg\ s^{-1}}$ will be above the dotted line and so we would expect a large majority to be \ngdd. This prediction is in agreement with Fig.~3 of \cite{lsg+15} which shows that only $\sim~15\%$ of the pulsars within this $\dot{E}$ range have $\gamma$-ray detections.

It was suggested by \cite{hsg+14} that pulsars are more difficult to detect in $\gamma$ rays when the $\gamma$-ray duty cycle is greater. However, taking $\Delta$ as a proxy for the extent of the $\gamma$-ray light curve, we can see from Fig.~\ref{FigWOfDelta} that easier expected $\gamma$-ray detection (i.e., lower $\Delta$) is associated with wider radio profiles. This implies that the lack of $\gamma$-rays in the \ngd sample, which typically have wide radio profiles (Fig.~\ref{FigWOfEdot}), is in general not due to a difficulty in the detection.

These observational results can all be explained by the following scenario: all, or at least most of, the radio pulsars in both samples emit $\gamma$ rays, given that they all have substantial values of $\dot{E}$. The \ngd pulsars are viewed too close to the rotation axis for $\gamma$ rays to be emitted along our line of sight, or such that the $\gamma$-ray light curve components are not sufficiently peaked to have been detected as pulsed emission. This is preferentially expected to happen if their magnetic inclination angle $\alpha$ is small, thereby explaining their wider radio profiles. The $\gamma$-ray beam is expected to decrease in size as $\dot{E}$ decreases (due to an increase in the thickness $w$ of the charge-depletion region), which is consistent with the slope of the boundary in radio pulse width between the two samples as a function of $\dot{E}$. Even if this scenario is not the dominant mechanism, it would still be expected to operate in conjunction with any other mechanism. A quantitative explanation of these results could potentially be used to constrain $\gamma$-ray emission models, the evolution of $w$ with $\dot E$ and radio beam models, although $\gamma$ ray detection biases would need to be considered.

This model of the $\alpha$-dependence of the $\gamma$-ray detectability is qualitatively similar to that recently proposed for millisecond pulsars by \cite{gt14} based on $\gamma$-ray light curve fitting, suggesting that the same model applies to the whole pulsar population. It is encouraging that very different types of analysis reach similar conclusions. It should also be stressed that our interpretation would work for both the two-pole caustic and outer gap models, demonstrating our conclusions are insensitive to the details of the $\gamma$-ray emission model. Additional evidence for this $\alpha$-dependence model is the correlation found between radio profile and $\gamma$-ray light curve morphologies. Light curves with the sharpest components and largest peak separations, suggestive of large $\alpha$, are indeed found to have narrower radio profiles.

There is overlap between the \grd and \ngd pulsars in ($W_{10}$, $\dot{E}$) space. This is expected since, besides $\alpha$, variations in parameters such as emission height or line of sight impact parameter will also affect the observed radio profile width. Several of the \ngd pulsars which occupy the region dominated by \grd pulsars are more distant than the majority of the \grd sample. It is therefore expected that some of these pulsars will be detected in $\gamma$ rays in the future as more data are acquired or instruments with greater sensitivity become available. Such detections could potentially then make the separation of the two samples in ($W_{10}$, $\dot{E}$) space even more pronounced.

There are some additional important consequences of these results. Firstly, the correlation between the radio and $\gamma$-ray light curve morphologies is consistent with models which place the $\gamma$-ray emission region high in the magnetosphere (such as the two-pole caustic and outer gap models), thereby adding more evidence that this type of model is at least qualitatively realistic. Additionally, the correlation of $W_{10}$ with $\dot{E}$ and $\gamma$-ray light curve morphology suggests that the radio beams, at least of these high-$\dot{E}$ pulsars, are largely filled. In the qualitative explanation of these correlations we have assumed that the radio profile width is directly related to the size of the open field line region. If the beams were too sparsely populated by patches of radio emission all correlations involving $W_{10}$ would weaken.

The results of this paper reinforce the conclusions of \cite{rwj15a}. This is first of all because in that paper a degree of radio beam symmetry was a key assumption for many of the analysed \grd pulsars, something implied by largely filled beams. Furthermore, we have demonstrated that $\gamma$-ray emission models which concentrate the $\gamma$-ray beam towards the rotational equator appear to be realistic. This confirms that a bias towards orthogonal rotators (larger $\alpha$) is expected for a \grd sample, making the trend of many \grd pulsars appearing to have low inclination angles, found by \cite{rwj15a}, even more peculiar. As discussed in \cite{rwj15b}, this implies that either pulsars are born with a relatively aligned magnetic field, or that the beam size depends on the inclination angle in a specific way. Both conclusions would affect the results of population synthesis. 

In this paper we have presented a new observational connection between the populations of young radio-loud $\gamma$-ray-detected and non-$\gamma$-ray-detected pulsars, namely that those with visible $\gamma$-ray emission have systematically narrower radio profiles. This connection can be interpreted in a model-independent fashion as a difference in the magnetic inclination angles of the two populations. Incorporating this new information into both synthesis codes and models of $\gamma$-ray emission will lead to an improved understanding of the $\gamma$-ray pulsar population and their particle acceleration and radiation mechanisms.

\section*{Acknowledgments}

We are very grateful to Geoff Wright for many helpful discussions. We would also like to thank the referee for constructive comments which helped improve this paper. The Parkes radio telescope is part of the Australia Telescope, which is funded by the Commonwealth Government for operation as a National Facility managed by CSIRO. Pulsar research at JBCA is supported by a Consolidated Grant from the UK Science and Technology Facilities Council (STFC).

\bibliographystyle{mn2e}
%\bibliographystyle{mnras}
%\bibliography{journals_apj,psrrefs} 

\begin{thebibliography}{}

\bibitem[\protect\citeauthoryear{{Abdo}, {Ackermann}, {Ajello}, {Baldini} \& et
  al.}{{Abdo} et~al.}{2010a}]{aaa+10a}
{Abdo} A.~A.,  {Ackermann} M.,  {Ajello} M.,  {Baldini} L.,    et al. 2010a,
  ApJ, 711, 64

\bibitem[\protect\citeauthoryear{{Abdo}, {Ajello}, {Allafort}, {Baldini} \& et
  al.}{{Abdo} et~al.}{2013}]{aaa+13}
{Abdo} A.~A.,  {Ajello} M.,  {Allafort} A.,  {Baldini} L.,    et al. 2013,
  ApJS, 208, 17

\bibitem[\protect\citeauthoryear{{Abdo}, {Ajello}, {Antolini}, {Baldini} \& et
  al.}{{Abdo} et~al.}{2010b}]{aaa+10c}
{Abdo} A.~A.,  {Ajello} M.,  {Antolini} E.,  {Baldini} L.,    et al. 2010b, ApJ,
  720, 26

\bibitem[\protect\citeauthoryear{{Barr}, {Champion}, {Kramer}, {Eatough} \& et
  al.}{{Barr} et~al.}{2013}]{bck+13}
{Barr} E.~D.,  {Champion} D.~J.,  {Kramer} M.,  {Eatough} R.~P.,    et al.
  2013, MNRAS, 435, 2234

\bibitem[\protect\citeauthoryear{{Bietenholz}, {Kondratiev}, {Ransom}, {Slane},
  {Bartel} \& {Buchner}}{{Bietenholz} et~al.}{2013}]{bkr+13}
{Bietenholz} M.~F.,  {Kondratiev} V.,  {Ransom} S.,  {Slane} P.,  {Bartel} N.,
    {Buchner} S.,  2013, MNRAS, 431, 2590

\bibitem[\protect\citeauthoryear{{Camilo}, {Ray}, {Ransom}, {Burgay} \& et
  al.}{{Camilo} et~al.}{2009}]{crr+09}
{Camilo} F.,  {Ray} P.~S.,  {Ransom} S.~M.,  {Burgay} M.,    et al. 2009, ApJ,
  705, 1

\bibitem[\protect\citeauthoryear{{Candy} \& {Blair}}{{Candy} \&
  {Blair}}{1986}]{cb86}
{Candy} B.~N.,  {Blair} D.~G.,  1986, ApJ, 307, 535

\bibitem[\protect\citeauthoryear{Cheng, Ho \& Ruderman}{Cheng
  et~al.}{1986}]{chr86a}
Cheng K.~S.,  Ho C.,    Ruderman M.,  1986, ApJ, 300, 500

\bibitem[\protect\citeauthoryear{{Craig}}{{Craig}}{2014}]{cra14}
{Craig} H.~A.,  2014, ApJ, 790, 102

\bibitem[\protect\citeauthoryear{{Dyks} \& {Rudak}}{{Dyks} \&
  {Rudak}}{2003}]{dr03}
{Dyks} J.,  {Rudak} B.,  2003, ApJ, 598, 1201

\bibitem[\protect\citeauthoryear{{Dyks}, {Rudak} \& {Demorest}}{{Dyks}
  et~al.}{2010}]{drd10}
{Dyks} J.,  {Rudak} B.,    {Demorest} P.,  2010, MNRAS, 401, 1781

\bibitem[\protect\citeauthoryear{Gil, Gronkowski \& Rudnicki}{Gil
  et~al.}{1984}]{ggr84}
Gil J.~A.,  Gronkowski P.,    Rudnicki W.,  1984, A\&A, 132, 312

\bibitem[\protect\citeauthoryear{Goldreich \& Julian}{Goldreich \&
  Julian}{1969}]{gj69}
Goldreich P.,  Julian W.~H.,  1969, ApJ, 157, 869

\bibitem[\protect\citeauthoryear{{Guillemot} \& {Tauris}}{{Guillemot} \&
  {Tauris}}{2014}]{gt14}
{Guillemot} L.,  {Tauris} T.~M.,  2014, MNRAS, 439, 2033

\bibitem[\protect\citeauthoryear{{Halpern}, {Camilo}, {Gotthelf}, {Helfand} \&
  et al.}{{Halpern} et~al.}{2001}]{hcg+01}
{Halpern} J.~P.,  {Camilo} F.,  {Gotthelf} E.~V.,  {Helfand} D.~J.,    et al.
  2001, ApJ, 552, L125

\bibitem[\protect\citeauthoryear{{Hou}, {Smith}, {Guillemot}, {Cheung} \& et
  al.}{{Hou} et~al.}{2014}]{hsg+14}
{Hou} X.,  {Smith} D.~A.,  {Guillemot} L.,  {Cheung} C.~C.,    et al. 2014,
  A\&A, 570, A44

\bibitem[\protect\citeauthoryear{{Johnston} \& {Weisberg}}{{Johnston} \&
  {Weisberg}}{2006}]{jw06}
{Johnston} S.,  {Weisberg} J.~M.,  2006, MNRAS, 368, 1856

\bibitem[\protect\citeauthoryear{{Keith}, {Eatough}, {Lyne}, {Kramer} \& et
  al.}{{Keith} et~al.}{2009}]{kel+09}
{Keith} M.~J.,  {Eatough} R.~P.,  {Lyne} A.~G.,  {Kramer} M.,    et al. 2009,
  MNRAS, 395, 837

\bibitem[\protect\citeauthoryear{Komesaroff}{Komesaroff}{1970}]{kom70}
Komesaroff M.~M.,  1970, Nature, 225, 612

\bibitem[\protect\citeauthoryear{{Laffon}, {Smith}, {Guillemot} \& {for the
  Fermi-LAT Collaboration}}{{Laffon} et~al.}{2015}]{lsg+15}
{Laffon} H.,  {Smith} D.~A.,  {Guillemot} L.,    {for the Fermi-LAT
  Collaboration} 2015, ArXiv e-prints, 1502.03251

\bibitem[\protect\citeauthoryear{{Li}, {Spitkovsky} \& {Tchekhovskoy}}{{Li}
  et~al.}{2012}]{lst12}
{Li} J.,  {Spitkovsky} A.,    {Tchekhovskoy} A.,  2012, ApJL, 746, L24

\bibitem[\protect\citeauthoryear{{Lorimer}, {Camilo} \& {McLaughlin}}{{Lorimer}
  et~al.}{2013}]{lcm13}
{Lorimer} D.~R.,  {Camilo} F.,    {McLaughlin} M.~A.,  2013, MNRAS, 434, 347

\bibitem[\protect\citeauthoryear{{Lorimer}, {Faulkner}, {Lyne}, {Manchester} \&
  et al.}{{Lorimer} et~al.}{2006}]{lfl+06}
{Lorimer} D.~R.,  {Faulkner} A.~J.,  {Lyne} A.~G.,  {Manchester} R.~N.,    et
  al. 2006, MNRAS, 372, 777

\bibitem[\protect\citeauthoryear{Lorimer \& Xilouris}{Lorimer \&
  Xilouris}{2000}]{lx00}
Lorimer D.~R.,  Xilouris K.~M.,  2000, ApJ, 545, 385

\bibitem[\protect\citeauthoryear{{Lyne}, {Graham-Smith}, {Weltevrede},
  {Jordan}, {Stappers}, {Bassa} \& {Kramer}}{{Lyne} et~al.}{2013}]{lgw+13}
{Lyne} A.,  {Graham-Smith} F.,  {Weltevrede} P.,  {Jordan} C.,  {Stappers} B.,
  {Bassa} C.,    {Kramer} M.,  2013, Science, 342, 598

\bibitem[\protect\citeauthoryear{Lyne \& Manchester}{Lyne \&
  Manchester}{1988}]{lm88}
Lyne A.~G.,  Manchester R.~N.,  1988, MNRAS, 234, 477

\bibitem[\protect\citeauthoryear{{Manchester}, {Hobbs}, {Teoh} \&
  {Hobbs}}{{Manchester} et~al.}{2005}]{mhth05}
{Manchester} R.~N.,  {Hobbs} G.~B.,  {Teoh} A.,    {Hobbs} M.,  2005, AJ, 129,
  1993

\bibitem[\protect\citeauthoryear{Michel}{Michel}{1987}]{mic87a}
Michel F.~C.,  1987, ApJ, 322, 822

\bibitem[\protect\citeauthoryear{Moffett \& Hankins}{Moffett \&
  Hankins}{1999}]{mh99}
Moffett D.~A.,  Hankins T.~H.,  1999, ApJ, 522, 1046

\bibitem[\protect\citeauthoryear{{Morris}, {Hobbs}, {Lyne}, {Stairs} \& et
  al.}{{Morris} et~al.}{2002}]{mhl+02}
{Morris} D.~J.,  {Hobbs} G.,  {Lyne} A.~G.,  {Stairs} I.~H.,    et al. 2002,
  MNRAS, 335, 275

\bibitem[\protect\citeauthoryear{{Muslimov} \& {Harding}}{{Muslimov} \&
  {Harding}}{2003}]{mh03}
{Muslimov} A.~G.,  {Harding} A.~K.,  2003, ApJ, 588, 430

\bibitem[\protect\citeauthoryear{{Pilia}, {Hessels}, {Stappers}, {Kondratiev},
  {Kramer} \& et al.}{{Pilia} et~al.}{2016}]{phs+15}
{Pilia} M.,  {Hessels} J.~W.~T.,  {Stappers} B.~W.,  {Kondratiev} V.~I.,
  {Kramer} M.,    et al. 2016, A\&A, 586, 92

\bibitem[\protect\citeauthoryear{Rankin}{Rankin}{1990}]{ran90}
Rankin J.~M.,  1990, ApJ, 352, 247

\bibitem[\protect\citeauthoryear{Rankin}{Rankin}{1993}]{ran93}
Rankin J.~M.,  1993, ApJ, 405, 285

\bibitem[\protect\citeauthoryear{{Ravi}, {Manchester} \& {Hobbs}}{{Ravi}
  et~al.}{2010}]{rmh10}
{Ravi} V.,  {Manchester} R.~N.,    {Hobbs} G.,  2010, ApJL, 716, L85

\bibitem[\protect\citeauthoryear{{Ray}, {Kerr}, {Parent}, {Abdo}, {Guillemot}
  \& et al.}{{Ray} et~al.}{2011}]{rkp+11}
{Ray} P.~S.,  {Kerr} M.,  {Parent} D.,  {Abdo} A.~A.,  {Guillemot} L.,    et
  al. 2011, ApJS, 194, 17

\bibitem[\protect\citeauthoryear{Romani}{Romani}{1996}]{rom96a}
Romani R.~W.,  1996, ApJ, 470, 469

\bibitem[\protect\citeauthoryear{{Romani} \& {Watters}}{{Romani} \&
  {Watters}}{2010}]{rw10}
{Romani} R.~W.,  {Watters} K.~P.,  2010, ApJ, 714, 810

\bibitem[\protect\citeauthoryear{Rookyard}{Rookyard}{2015}]{roo15}
Rookyard S.~C.,  2015, PhD thesis, The University of Manchester

\bibitem[\protect\citeauthoryear{{Rookyard}, {Weltevrede} \&
  {Johnston}}{{Rookyard} et~al.}{2015a}]{rwj15a}
{Rookyard} S.~C.,  {Weltevrede} P.,    {Johnston} S.,  2015a, MNRAS, 446, 3367

\bibitem[\protect\citeauthoryear{{Rookyard}, {Weltevrede} \&
  {Johnston}}{{Rookyard} et~al.}{2015b}]{rwj15b}
{Rookyard} S.~C.,  {Weltevrede} P.,    {Johnston} S.,  2015b, MNRAS, 446, 3356

\bibitem[\protect\citeauthoryear{{Smith}, {Guillemot}, {Camilo}, {Cognard} \&
  et al.}{{Smith} et~al.}{2008}]{sgc+08a}
{Smith} D.~A.,  {Guillemot} L.,  {Camilo} F.,  {Cognard} I.,    et al. 2008,
  A\&A, 492, 923

\bibitem[\protect\citeauthoryear{{Spitkovsky}}{{Spitkovsky}}{2006}]{spi06}
{Spitkovsky} A.,  2006, ApJ, 648, L51

\bibitem[\protect\citeauthoryear{Tauris \& Manchester}{Tauris \&
  Manchester}{1998}]{tm98}
Tauris T.~M.,  Manchester R.~N.,  1998, MNRAS, 298, 625

\bibitem[\protect\citeauthoryear{Taylor \& Cordes}{Taylor \&
  Cordes}{1993}]{tc93}
Taylor J.~H.,  Cordes J.~M.,  1993, ApJ, 411, 674

\bibitem[\protect\citeauthoryear{{Theureau}, {Parent}, {Cognard}, {Desvignes},
  {Smith} \& et al.}{{Theureau} et~al.}{2011}]{tpc+11}
{Theureau} G.,  {Parent} D.,  {Cognard} I.,  {Desvignes} G.,  {Smith} D.~A.,
  et al. 2011, A\&A, 525, A94

\bibitem[\protect\citeauthoryear{{van Leeuwen}, {Kasian}, {Stairs}, {Lorimer},
  {Camilo}, {Chatterjee} \& et al.}{{van Leeuwen} et~al.}{2015}]{vks+15}
{van Leeuwen} J.,  {Kasian} L.,  {Stairs} I.~H.,  {Lorimer} D.~R.,  {Camilo}
  F.,  {Chatterjee} S.,    et al. 2015, ApJ, 798, 118

\bibitem[\protect\citeauthoryear{{Wang}, {Pi}, {Zheng}, {Deng}, {Wen}, {Ye},
  {Guan}, {Liu} \& {Xu}}{{Wang} et~al.}{2014}]{wpz+14}
{Wang} H.~G.,  {Pi} F.~P.,  {Zheng} X.~P.,  {Deng} C.~L.,  {Wen} S.~Q.,  {Ye}
  F.,  {Guan} K.~Y.,  {Liu} Y.,    {Xu} L.~Q.,  2014, ApJ, 789, 73

\bibitem[\protect\citeauthoryear{{Watters}, {Romani}, {Weltevrede} \&
  {Johnston}}{{Watters} et~al.}{2009}]{wrw+09}
{Watters} K.~P.,  {Romani} R.~W.,  {Weltevrede} P.,    {Johnston} S.,  2009,
  ApJ, 695, 1289

\bibitem[\protect\citeauthoryear{{Weltevrede} \& {Johnston}}{{Weltevrede} \&
  {Johnston}}{2008}]{wj08b}
{Weltevrede} P.,  {Johnston} S.,  2008, MNRAS, 391, 1210

\bibitem[\protect\citeauthoryear{{Weltevrede}, {Johnston}, {Manchester}, {Bhat}
  \& et al.}{{Weltevrede} et~al.}{2010}]{wjm+10}
{Weltevrede} P.,  {Johnston} S.,  {Manchester} R.~N.,  {Bhat} R.,    et al.
  2010, Proc. Astr. Soc. Aust., 27, 64

\end{thebibliography}

\bsp

\onecolumn
\appendix
\section{Classification of $\gamma$-ray light curves}

This appendix briefly summarises our classification of the $\gamma$-ray light curve for each \grd pulsar, from which we allocated the components as either ``broad'' or ``narrow'' (see $\S$~\ref{SectWidthVersusMorphology}). In all cases the phases of the components are given relative to the maximum in the radio profile. All figures used can be found in the online supplement to the Second \emph{Fermi} Catalogue \citep{aaa+13} at http://fermi.gsfc.nasa.gov/ssc/data/access/lat/2nd\_PSR\_catalog/combined\_2PC\_lightcurves.pdf unless otherwise stated.\\
\\
\emph{J0205+6449}: Two components at phases $\sim0.075$ and $\sim0.575$, both narrower than 0.15$P$.\\
\emph{J0534+2200}: Two components at phases $\sim0.1$ and $\sim0.5$, both narrower than 0.15$P$.\\
\emph{J0631+1036}: A single component at phase $\sim0.5$, broader than 0.15$P$\\
\emph{J0729$-$1448}: A single component at phase $\sim0.6$, narrower than 0.15$P$.\\
\emph{J0742$-$2822}: A single component at phase $\sim0.6$, narrower than 0.15$P$.\\
\emph{J0835$-$4510 (Vela)}: Two components at phases $\sim0.15$ and $\sim0.55$, both narrower than 0.15$P$, and a lower-intensity partially resolved component at phase $\sim0.25$ of uncertain width at half-maximum intensity.\\
\emph{J0908$-$4913}: Two components at phases $\sim0.1$ and $\sim0.6$, both narrower than 0.15$P$.\\
\emph{J0940$-$5428}: A single component at phase $\sim0.45$, broader than 0.15$P$.\\
\emph{J1016$-$5857}: Two components at phases $\sim0.15$ and $\sim0.55$, both narrower than 0.15$P$, and a lower-intensity component at phase $\sim0.2$, having a full width at half maximum $\sim0.15P$.\\
\emph{J1028$-$5819}: Two components at phases $\sim0.2$ and $\sim0.65$, both narrower than 0.15$P$.\\
\emph{J1048$-$5832}: Two components at phases $\sim0.1$ and $\sim0.55$, both narrower than 0.15$P$.\\
\emph{J1055$-$6028}: Fig.~1 of \cite{hsg+14} shows three components at phases $\sim0.15$, $\sim0.45$ and $\sim0.8$, the former two broader than 0.15$P$ and the latter narrower than 0.15$P$.\\
\emph{J1105$-$6107}: Two components at phases $\sim0.1$ and $\sim0.6$, both narrower than 0.15$P$.\\
\emph{J1119$-$6127}: One component at phase $\sim0.45$, broader than 0.15$P$, and a slightly lower-intensity partially-resolved component at phase $\sim0.3$ with a width at half maximum intensity which is uncertain although potentially narrower than 0.15$P$.\\
\emph{J1124$-$5916}: Two components at phases $\sim0.15$ and $\sim0.65$, both narrower than 0.15$P$.\\
\emph{J1357$-$6429}: A single component at phase $\sim0.35$, broader than 0.15$P$.\\
\emph{J1420$-$6048}: One component at phase $\sim0.5$, broader than 0.15$P$, and a lower-intensity component at phase $\sim0.2$ with a width at half maximum intensity which is uncertain although probably narrower than 0.15$P$.\\
\emph{J1509$-$5850}: Two components at phases $\sim0.25$ and $\sim0.55$, both broader than 0.15$P$.\\
\emph{J1513$-$5908}: A single component at phase $\sim0.35$, broader than 0.15$P$.\\
\emph{J1531$-$5610}: A single component at phase $\sim0.4$, having a full width at half maximum intensity $\sim0.15P$. We designate this as a ``narrow'' component, but note that it is marginal. We also note that the radio profile width ($34.6\degree$) places this pulsar close to the transition from ``narrow'' to ``broad'' single-component light curves in Fig.~\ref{FigWOfDelta}, reinforcing the idea that this may be a borderline case between the two classes.\\
\emph{J1709$-$4429}: Two components at phases $\sim0.25$ and $\sim0.55$, both broader than 0.15$P$.\\
\emph{J1718$-$3825}: A single component at phase $\sim0.4$, broader than 0.15$P$.\\
\emph{J1739$-$3023}: Fig.~2 of \cite{lsg+15} shows two components at phases $\sim0.1$ and $\sim0.5$, both broader than 0.15$P$.\\
\emph{J1801$-$2451}: Two components at phases $\sim0.05$ and $\sim0.55$, both narrower than 0.15$P$.\\
\emph{J1831$-$0952}: Fig.~2 of \cite{lsg+15} shows two components at phases $\sim0.35$ and $\sim0.5$, both broader than 0.15$P$.\\
\emph{J1833$-$1034}: Two components at phases $\sim0.15$ and $\sim0.6$, both narrower than 0.15$P$.\\
\emph{J1835$-$1106}: Two components at phases $\sim0.15$ and $\sim0.5$, both broader than 0.15$P$.\\
\emph{J1856+0113}: Fig.~2 of \cite{lsg+15} shows two components at phases $\sim0.1$ and $\sim0.6$, both narrower than 0.15$P$.\\
\emph{J1857+0143}: Fig.~2 of \cite{lsg+15} shows a single component at phase $\sim0.3$, broader than 0.15$P$.\\
\emph{J1907+0602}: Two components at phases $\sim0.2$ and $\sim0.6$, both narrower than 0.15$P$.\\
\emph{J1913+0904}: Fig.~1 of \cite{hsg+14} shows three components at phases $\sim0.3$, $\sim0.65$ and $\sim0.95$, the former two broader than 0.15$P$ and the latter narrower than 0.15$P$.\\
\emph{J1952+3252}: Two components at phases $\sim0.15$ and $\sim0.65$, both narrower than 0.15$P$.\\
\emph{J2032+4127}: Two components at phases $\sim0.1$ and $\sim0.6$, both narrower than 0.15$P$.\\
\emph{J2229+6114}: One dominant component at phase $\sim0.45$, broader than 0.15$P$, and one lower-intensity leading component apparent only at lower $\gamma$-ray energies (0.1 - 0.3 GeV) which \cite{aaa+13} used to determine $\Delta$.\\
\emph{J2240+5832}: Two components at phases $\sim0.1$ and $\sim0.6$, both narrower than 0.15$P$.\\

\label{lastpage}

\end{document}